\documentstyle[aps,epsf,floats,prl]{revtex}
\begin{document}
%
\twocolumn[\hsize\textwidth\columnwidth\hsize\csname @twocolumnfalse\endcsname
\title{Structure and Stability of Si(114)-${\bf (2\times 1)}$}
\author{S.C. Erwin,\cite{corresponding} A.A. Baski, and L.J. Whitman}
\address{Naval Research Laboratory, Washington DC 20375}
\date{\today}
\maketitle
\begin{abstract}
We describe a recently discovered stable planar surface of silicon,
Si(114).  This high-index surface, oriented 19.5$^\circ$ away from
(001) toward (111), undergoes a $2\times 1$ reconstruction.  We
propose a complete model for the reconstructed surface based on
scanning tunneling microscopy images and first-principles total-energy
calculations.  The structure and stability of Si(114)-$(2\times 1)$
arises from a balance between surface dangling bond reduction and
surface stress relief, and provides a key to understanding the
morphology of a family of surfaces oriented between (001) and (114).\\
\mbox{}\\
(A preprint with high-resolution figures is at 
{\tt http://cst-www.nrl.navy.mil/papers/si114.ps})
\end{abstract}
%
]
The (001) and (111) surfaces of silicon are its primary growth and
cleavage planes, respectively.  Surfaces with orientations between
these two planes are both intrinsically interesting and
technologically important.  For these intermediate orientations,
surface morphology depends strongly on the orientation angle.  For
example, small misorientations away from the low-index planes produce
vicinal surfaces consisting of terraces separated by steps
\cite{swartzentruber,alerhand,williams}, whereas larger angles often
lead to sawtooth-like, grooved surfaces \cite{baski95a,song95}.
Between (001) and (111), only two orientations are known to form
planar surfaces with stable reconstructions: Si(113)
\cite{knall,dabrowski} and Si(5\,5\,12) \cite{baski95b}.  Recently,
(114) surfaces have been observed both on cylindrical Si samples
\cite{suzuki} and within etch pits formed on Si(001) \cite{kendall}.
In addition, sawtooth-like structures composed of (114) and (113)
facets have been observed on samples oriented between (114) and (113)
\cite{song95,suzuki}.  These results suggest that Si(114) is a 
planar surface that is thermodynamically stable against
faceting.

\begin{figure}[p]
\epsfxsize=5.1cm\centerline{\epsffile{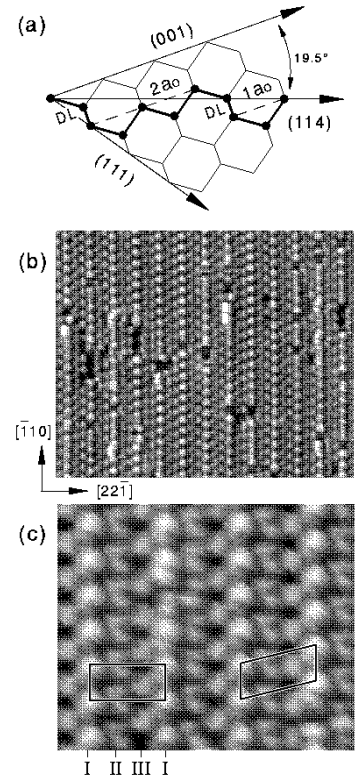}}
\caption{(a) Side view of the Si crystal lattice, showing one unit
cell of the bulk-terminated (114) surface. Dashed lines emphasize the
alternating single- and double-width (001) terraces separated by
double-layer (DL) steps. (b) Filled-state STM gray-scale image
($210\times 180$ \AA) of the clean Si(114) surface. (c)
Atomic-resolution, filled-state image ($65\times 54$ \AA) highlighting
the three types of observed row structures.  Unit cells are drawn to
indicate two types of local symmetry: $2\times 1$ [left] and
$c(2\times 2)$ [right]. The $2\times 1$ unit cell is $16.3\times 7.7$ \AA.}
\end{figure}

The (114) plane can be considered a highly vicinal (001) surface,
oriented $\theta$=19.5$^\circ$ from (001) toward (111).  The structure
of the ideal bulk-terminated (114) surface is an alternating sequence
of one- and two-unit-cell wide (001) terraces separated by
double-layer steps, as shown in Fig.~1(a).  This basic
terrace-plus-step morphology also describes real surfaces over a
large range of orientation angles: from slightly vicinal (001)
surfaces with large variable-width terraces to steeper orientations
consisting of three-unit-cell wide terraces separated by double-layer
steps ($\theta$=11.4$^\circ$) \cite{hanbuecken}. It is interesting to
ask at what orientation angle this morphology breaks down, and for
what reasons. For example, the Si(113) surface ($\theta$=25.2$^\circ$)
has a stable $3\times 2$ reconstruction that cannot be simply
described as the expected sequence of one-unit-cell terraces and
double-layer steps \cite{dabrowski}.  For a surface with (114)
orientation, one is therefore led to consider three possibilities: (i)
a periodic surface with a terrace-plus-step morphology; (ii) a
periodic surface without a terrace-plus-step morphology, similar
perhaps to Si(113)-$(3\times 2)$; or (iii) a non-periodic
(e.g.~grooved) surface.  In this Letter we use scanning tunneling
microscopy (STM) to definitively establish that Si(114)-$(2\times 1)$
is a planar surface with a stable terrace-plus-step reconstruction.
We propose a complete structural model for the reconstructed surface,
and use first-principles electronic-structure methods to provide
strong theoretical evidence in support of this model.  The structure
of Si(114) reveals it to be a natural extension of vicinal (001);
indeed, it marks the endpoint of a family of surfaces having closely
related structure.  

The clean Si(114) surface \cite{experiments} appears in
constant-current STM images as a well-ordered periodic array of row
structures oriented along the [$\overline{1}$10] direction
[Fig.~1(b)].  The period of this structure in the [22$\overline{1}$]
direction is approximately 16 \AA, equal to the length of the
bulk-terminated (114) unit cell.  Along each row, the period in the
[$\overline{1}$10] direction is 7.7 \AA, twice that of the
bulk-terminated unit cell.  Atomically resolved images reveal three
distinct types of row structures, labeled I, II, and III in Fig.~1(c).
The phase relationship along [$\overline{1}$10] between these row
structures varies over different regions of the surface, giving rise
to areas of local $2\times 1$ as well as $c(2\times 2)$ periodicity
[see Fig.~1(c)].

Our proposed structural model for Si(114)-$(2\times 1)$ is closely
related to the structure of vicinal Si(001).  The Si(001) surface
undergoes a reconstruction in which the surface atoms pair up and bond
to reduce the number of surface dangling bonds, forming parallel rows
of dimers.  For vicinal surfaces with $\theta$ greater than
4--5$^\circ$, the great majority of resulting (001) terraces consist
of dimers oriented parallel to the step edges (B-type terraces), with
these B-type terraces separated by double-layer steps ($D_B$ steps)
\cite{swartzentruber}.  Chadi first proposed that $D_B$ steps are
rebonded, i.e. they incorporate another row of atoms at the step edge
to reduce the number of dangling bonds \cite{chadi}.  This basic
morphology---B-type terraces separated by rebonded $D_B$ steps---has
also been observed in STM studies of curved Si samples having regions
with local orientation $\theta\sim$ 9--11$^\circ$, where the (001)
terrace width is only three or four unit cells \cite{hanbuecken}. We
propose that a variant of this basic morphology also describes the
very narrow terraces found on Si(114).  Specifically, our structural
model for Si(114)-$(2\times 1)$, shown in Fig.~2, is an alternating
sequence of single- and double-width B-type (001) terraces, separated
by rebonded $D_B$ steps (stepping up from the double-width terrace)
and non-rebonded $D_B$ steps (stepping down from the double-width
terrace). The presence of both rebonded and non-rebonded steps
provides an important mechanism for stress relief within the unit
cell, as discussed in detail below.

This structural model is strongly supported by the results of
first-principles calculations of equilibrium surface geometry and
surface energy, and the resulting simulated STM images.  The
calculations were performed in a supercell slab geometry (inversion
symmetric) with ten layers of Si and a vacuum layer equivalent to six
Si layers.  Full structural relaxation was performed using the Corning
electronic-structure code of Allan, Teter, and Payne \cite{allan},
which solves the Kohn-Sham equations in the local-density
approximation (LDA) with a plane-wave basis and norm-conserving
pseudopotentials \cite{teter}.  The kinetic-energy cutoff was 10 Ry,
and a single $k$-point was used throughout.  Structural relaxation was
performed on all the atoms until the rms force was less than 0.1
eV/\AA \cite{convergence}.

\begin{figure}[t]
\epsfysize=8.3cm\centerline{\epsffile{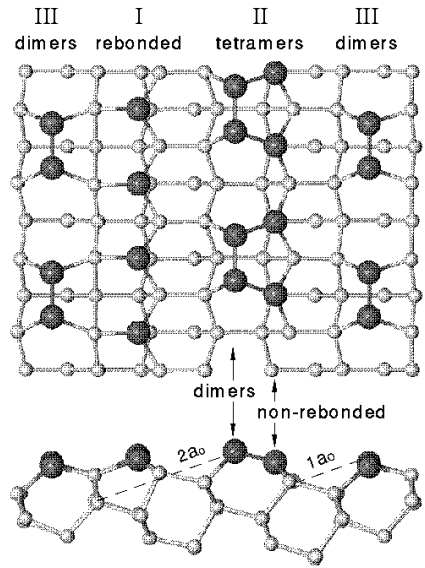}}
\caption{Top and side views of proposed model for Si(114)-$2(\times
1)$ reconstruction, with fully relaxed geometry. Labeled structures
correspond to the STM features seen in Figs.~1 and 3. A dimer plus two
non-rebonded step-edge atoms is called a ``tetramer.''}
\end{figure}

The fully relaxed surface geometry for our proposed model includes
three prominent row structures per unit cell, corresponding to the
three types of rows seen in the STM images.  Row I in the images
arises from the line of rebonding atoms at the rebonded $D_B$ step
edge; row II arises from the row of tetramers (i.e. the combination of
the dimers on the double-width terrace and the non-rebonded $D_B$ step-edge
atoms); and row III arises from the dimers on the single-width terrace.
The correspondence between theory and experiment is examined in detail
in Fig.~3. Figs.~3(a) and (b) show STM images for filled and empty
states, respectively, from the same surface area; Figs.~3(c) and (d)
show numerically simulated STM images (constant current surfaces)
calculated by integrating the local-state density over filled and
empty states, respectively, near the Fermi level.  In the filled-state
STM image [Fig.~3(a)], rows I, II, and III all show 2$a_0$ periodicity
along [$\overline{1}$10] ($a_0$=3.84 \AA~is the surface lattice
constant): rows I and III each appear as an array of maxima with
2$a_0$ spacing, whereas row II has a more variable appearance.  In the
empty-state image [Fig.~3(b)], all three rows also exhibit 2$a_0$
periodicity, but the features are more diffuse within each row.

\begin{figure}[t]
\epsfysize=7.5cm\centerline{\epsffile{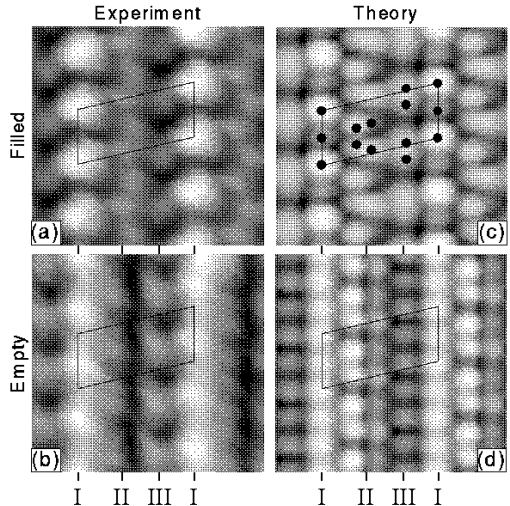}}
\caption{(Left) STM images ($33\times 31$ \AA) of Si(114) measured
for: (a) filled states, 1.2 V; (b) empty states, $-$1.7 V V. (Right)
Simulated STM images calculated for: (c) filled states; (d) empty
states. Black dots indicate projected atom positions.  The
simulated images are for a surface with $c(2\times 2)$ symmetry.}
\end{figure}

The general appearance of the rows is accurately reproduced in the
simulated STM images based on our model. In the simulated filled-state
image [Fig.~3(c)], row I is most prominent, and the expected 2$a_0$
periodicity is apparent in all three rows, in good agreement with the
STM data.  Along row I, weak maxima are visible between the dominant
peaks spaced 2$a_0$ apart; these maxima are only occasionally resolved
in STM images.  Row II has a zigzag-like internal structure,
consistent with the variable appearance observed experimentally.  In
the simulated empty-state image [Fig.~3(d)], the 2$a_0$ periodicity is
still evident in row II, but is much weaker in rows I and III, as
observed in the experimental image. Finally, the experimental
observation of areas with local $2\times 1$ and $c(2\times 2)$
symmetry is also consistent with our model.  We have calculated total
energies of both reconstructions and find differences only at the
level of 1--2 meV/\AA$^2$.

Several of the topographic features described above are a consequence
of a structural distortion that occurs within each row. For example,
at equilibrium the dimers that constitute row III are each tilted
(buckled) so that the two dimer atoms have a theoretical height
difference of $\Delta z$=0.17 \AA.  This buckling occurs on flat
Si(001) as well, and is energetically favored by the transfer of an
electron from the lower to upper atom of the dimer.  A similar
buckling occurs along the line of rebonding atoms that constitute row
I ($\Delta z$=0.18 \AA).  Both of these bucklings, although small,
give rise to a large corrugation of the constant-current surfaces and
are critical to explaining the appearance of rows I and III as lines
of maxima with 2$a_0$ spacing. The tetramers, which constitute row II,
have a more complicated geometry that gives rise to the internal
structure visible in both the experimental and simulated images.  In
the equilibrium surface geometry shown in Fig.~2, the tetramers are
essentially symmetric (no buckling). In the course of relaxing from
slightly different initial geometries, however, we found a variation
on this structure (also at equilibrium and with essentially the same
total energy) with a highly buckled tetramer ($\Delta z$=0.30 and 0.15
\AA~for the dimer pair and step-edge atom pair, respectively).  The
simulated images in Fig.~3 are from a surface with buckled tetramers;
the similar total energies suggest the possibility of different
metastable geometries within row II, explaining its variable
appearance in the experimental images.

Experimentally, we observe that Si wafers oriented within 0.5$^\circ$
of (114) form large terraces of (114) separated by steps.  Wafers with
orientations a few degrees off (114) toward (111) form grooved
surfaces consisting of (114) and (113) facets \cite{baski96},
consistent with earlier x-ray scattering results \cite{song95} and {\it
ex-situ} atomic force microscopy \cite{song95}.  These observations
suggest that the surface energy of (114) is quite low, perhaps
comparable to Si(113)-$(3\times 2)$.  We have investigated the
stability of our Si(114)-$(2\times 1)$ reconstruction model, relative
to low-index reconstructed surfaces, by calculating surface energies,
$E_s= [E_t(N) - NE_t^{\rm bulk}]/2$. Here $E_t(N)$ is the total energy
of the supercell containing $N$ atoms, $E_t^{\rm bulk}$ is the total
energy per atom of bulk Si (computed using the same lattice vectors,
energy cutoff, and zone sampling as for the surface calculations), and
the factor of two accounts for the two surfaces per unit cell. We find
the surface energies for the fully relaxed $2\times 1$ reconstructions
of Si(111), (001), and (114) to be 84, 86, and 85 meV/\AA$^2$,
respectively, indicating that Si(114)-$(2\times 1)$ is approximately
as stable as the low-index (111)-$(2\times 1)$ and (001)-$(2\times 1)$
surfaces.

While our proposed model---a terraced surface with $D_B$ steps
alternately rebonded and non-rebonded---was guided by our STM results,
we have also carried out fully relaxed total-energy calculations for a
number of alternative structures.  All lead to simulated STM images in
strong disagreement with experiment, and most lead to significantly
higher surface energies than our proposed model. Two of these models,
however, shed light on the mechanism that ultimately stabilizes
Si(114): (Model A) a terraced surface with every $D_B$ step rebonded;
and (Model B) a terraced surface with every $D_B$ step
non-rebonded. Because rebonded steps have fewer surface dangling bonds
than non-rebonded steps, one might naively predict Model A to be
strongly favored relative to both Model B and our model.  When fully
relaxed, however, we find that Model A has a surface energy of 87
meV/\AA$^2$, very slightly higher than our model (Model B is
significantly higher, at 97 meV/\AA$^2$). 

We believe
that the alternation of rebonded and non-rebonded step edges is
ultimately favored because it leads to an optimal balance between
surface dangling bond reduction and surface stress relief.  
For isolated $D_B$ steps
on Si(001), rebonded steps are energetically preferred to non-rebonded
steps because the number of dangling bonds is reduced.  However,
rebonded $D_B$ steps contribute a tensile stress in the direction
perpendicular to the step edge.  For sufficiently wide terraces, this
stress is elastically relieved, so that rebonded steps are favored.
For very short terraces, however, this relief mechanism is not
available, and the resulting energy penalty may be larger than the
energy gain from dangling-bond reduction.  {\em Si(114) balances these
competing effects by incorporating both a rebonded and a non-rebonded
$D_B$ step in each unit cell, allowing for both dangling-bond
reduction and stress relief within the unit cell.}

This mechanism of stress relief can be demonstrated quantitatively by
calculating the LDA surface stress tensor for each model,
$\sigma_{ij}=A^{-1} dE_s/d\epsilon_{ij}$, where $\epsilon_{ij}$ is the
surface strain tensor and $A$ is the surface cell area \cite{meade}.
Based on the above arguments, for Model A one expects a large tensile
stress in the $x$-direction (i.e. perpendicular to the step edge),
whereas for Model B and our model one expects a smaller (compressive or
tensile) stress.  Indeed, we find the values of $\sigma_{xx}$ to
be 0.27 eV/\AA$^2$ (tensile) for Model A, $-$0.03 eV/\AA$^2$
(compressive) for Model B, and 0.11 eV/\AA$^2$ for our
model.  These numerical results are consistent with the qualitative
scenario described above: Model A has a low density of dangling bonds
but high surface stress, which incurs a large energy penalty; Model B
has a low surface stress but high density of dangling bonds, which
also incurs a large energy penalty.  Our proposed model compromises on
both dangling bonds and surface stress, and by so doing is
the most favorable.

The competition between the energy gain from dangling-bond reduction
and the energy penalty from tensile stress at rebonded $D_B$ steps
provides a framework for understanding the surface morphology of a
surprisingly large class of orientations, from slightly vicinal
Si(001) up to Si(114) ($\theta$=19.5$^\circ$).  For orientations up to
(114), the stress at rebonded $D_B$ steps can be relieved either on
the terraces or by the presence of non-rebonded steps, and
consequently the terrace-plus-step morphology is favored.  For
example, curved surfaces with local (119) and (117) orientations
($\theta$=8.9$^\circ$ and 11.4$^\circ$, respectively) form
well-ordered B-type terraces separated exclusively by rebonded $D_B$
step edges \cite{hanbuecken,adams}.  Moreover, planar surfaces with
(115) orientation ($\theta$=15.8$^\circ$) consist of a complicated
arrangement of B-type terraces separated by both rebonded
and non-rebonded steps \cite{baski96}.  Beyond (114), however, the
terraces become too narrow to afford sufficient stress relief, and so
the terrace-plus-step morphology is no longer favored, leading to
fundamentally different surface morphologies. For example, surfaces
with orientations between (114) and (113) form mesoscale sawtooth-like
grooves \cite{song95,suzuki}, and surfaces with (113) orientation are
stabilized by a more complex $3\times 2$ reconstruction
\cite{knall,dabrowski}.  We therefore conclude that Si(114) marks the
endpoint of a family of orientations, each member of which has a
related surface morphology consisting of (001) terraces separated by
$D_B$ steps.

Computational work was supported by the Cornell Theory Center
and by a grant of HPC time from the DoD Shared Resource
Center MAUI. Some computational results were obtained using 
the software program PlaneWave (Biosym Technologies).
This work was funded by ONR and an NRL/NRC postdoctoral fellowship (AAB).
%
%

%
%
%
%


\begin{references}
\bibitem{experiments} The experiments were performed in ultrahigh
vacuum (UHV) with Si wafers oriented to within 0.5$^\circ$ of (114).
The chemically precleaned samples were mounted on a button heater in
UHV, flashed to $\sim$1150 $^\circ$C for 60 s (pressure $< 2 \times
10^{-9}$ Torr), and then slowly cooled (see Ref.~\cite{baski95a}).
This procedure resulted in contaminant-free surfaces with
well-ordered terraces greater than 0.1 $\mu$m $\times$ 0.1 $\mu$m
(determined only by the wafer miscut).
\end{references}
\end{document}